\documentclass{article}
\usepackage[utf8]{inputenc}
\usepackage{cite,fullpage}
\usepackage{url}
\usepackage{csquotes}
\usepackage{psfrag}
\usepackage{amsmath,amsfonts,amssymb}
\usepackage[dvips]{graphicx}
\usepackage{color}
\usepackage{wrapfig}
\usepackage{indentfirst}
\usepackage{epstopdf}
\usepackage{hyperref}
\usepackage{caption}
\usepackage{subcaption}
\usepackage{pgf}

\title{Data-driven modeling for different stages of pandemic response\footnote{\MakeLowercase{\MakeUppercase{T}o appear in the  ``\MakeUppercase{J}ournal of the \MakeUppercase{I}ndian \MakeUppercase{I}nstitute of \MakeUppercase{S}cience," \MakeUppercase{V}olume 100.}}}
\author{Aniruddha Adiga$^1$, \  Jiangzhuo Chen$^1$, Madhav Marathe$^{1,3}$ \\ Henning Mortveit$^{1,2}$,  \ Srinivasan Venkatramanan$^1$ \ and \ Anil Vullikanti$^{1,3}$ \\
$^1$Biocomplexity Institute and Inititiative \\
$^2$Department of Systems Engineering and Environment \\
$^3$Department of Computer Science\\
University of Virginia
}
\date{}

\begin{document}

\maketitle

\begin{abstract}
Some of the key questions of interest during the COVID-19 pandemic (and all outbreaks) include: where did the disease start, how is it spreading, who is at risk, and how to control the spread. There are a large number of complex factors driving the spread of pandemics, and, as a result, multiple modeling techniques play an increasingly important role in shaping public policy and decision making. As different countries and regions go through phases of the pandemic, the questions and data availability also changes.  Especially of interest is aligning model development and data collection to support response efforts at each stage of the pandemic. The COVID-19 pandemic has been unprecedented in terms of real-time collection and dissemination of a number of diverse datasets, ranging from disease outcomes, to mobility, behaviors, and socio-economic factors. 
The data sets have been critical from the perspective of disease modeling and analytics to support policymakers in real-time. In this overview article, we survey the data landscape around COVID-19, with a focus on how such datasets have aided modeling and response through different stages so far in the pandemic. We also discuss some of the current challenges  and the needs that will arise as we plan our way 
out of the pandemic. 
\end{abstract}

\section{Introduction}
\label{sec:intro}
\noindent
As the SARS-CoV-2 pandemic has demonstrated, the spread of a highly infectious disease is a complex dynamical process. 
A large number of factors are at play as infectious diseases spread, including variable individual susceptibility to the pathogen (e.g., by age and health conditions), variable individual behaviors (e.g., compliance with social distancing and the use of masks), differing response strategies implemented by governments (e.g., school and workplace closure policies and criteria for testing), and potential availability of pharmaceutical interventions. 
Governments have been forced to respond to the rapidly changing dynamics of the pandemic, and are becoming increasingly reliant on different modeling and analytical techniques to understand, forecast, plan and respond; this includes statistical methods and decision support methods using multi-agent models, such as: ($i$) forecasting epidemic outcomes (e.g., case counts, mortality and hospital demands), using a diverse set of data-driven methods e.g., ARIMA type time series forecasting, Bayesian techniques and deep learning, e.g.,~\cite{adhikari:kdd19, perone:ssrn20, desai:hs19, Reich2019AccuracyOR, funk:epidemics18}, 
($ii$) disease surveillance, e.g., ~\cite{healthmap, fung:wpsar}, and ($iii$) counter-factual analysis of epidemics using multi-agent models, e.g.,~\cite{chinazzi:science20, Britton2020, rocklov2020covid, ferguson2020report, eubank:nature04, marathe:cacm13}; indeed, the results of~\cite{covid2020forecasting, ferguson2020report} were very influential in the early decisions for lockdowns in a number of countries. 

\noindent
The specific questions of interest change with the stage of the pandemic. In the \emph{pre-pandemic} stage, the focus was on understanding how the outbreak started, epidemic parameters, and the risk of importation to different regions. Once outbreaks started-- the \emph{acceleration} stage, the focus is on determining the growth rates, the differences in spatio-temporal characteristics, and testing bias. In the \emph{mitigation} stage, the questions are focused on non-prophylactic interventions, such as school and work place closures and other social-distancing strategies, determining the demand for healthcare resources, and testing and tracing. In the \emph{suppression} stage, the focus shifts to using prophylactic interventions, combined with better tracing. 
These phases are not linear, and overlap with each other. For instance, the acceleration and mitigation stages of the pandemic might overlap
spatially, temporally as well as within certain social groups.

\noindent
Different kinds of models are appropriate at different stages, and for addressing different kinds of questions. For instance, statistical and machine learning models are very useful in forecasting and short term projections. However, they are not very effective for longer-term projections, understanding the effects of different kinds of interventions, and counter-factual analysis. Mechanistic models are very useful for such questions. Simple compartmental type models, and their extensions, namely, structured metapopulation models are useful for several population level questions. However, once the outbreak has spread, and complex individual and community level behaviors are at play, multi-agent  models are most effective, since they allow for a more systematic representation of complex social interactions, individual and
collective behavioral adaptation and public policies.

\noindent
As with any mathematical modeling effort, data plays a big role in 
the utility of such models. Till recently, data on infectious diseases was very hard to obtain due to various issues, such as privacy and sensitivity of the data (since it is information about individual health), and logistics of collecting such data. The data landscape during the SARS-CoV-2 pandemic has been very different: a large number of datasets are becoming available, ranging from disease outcomes (e.g., time series of the number of confirmed cases, deaths, and hospitalizations), some characteristics of their locations and demographics, healthcare infrastructure capacity (e.g., number of ICU beds, number of healthcare personnel, and ventilators), and various kinds of behaviors (e.g., level of social distancing, usage of PPEs); see \cite{alamo2020open, alamo2020data, shuja2020covid} for comprehensive surveys on available datasets.

\noindent
However, using these datasets for developing good models, and addressing important public health questions remains challenging. The goal of this article is to use the widely accepted stages of a pandemic as a guiding framework to highlight a few important problems that require attention in each of these stages. We will aim to provide a succinct model-agnostic formulation  while identifying the key datasets needed, how they can be used, and the challenges arising in that process. We will also use SARS-CoV-2 as a case study unfolding in real-time, and highlight some interesting peer-reviewed and preprint literature that pertains to each of these problems. An important point to note is the necessity of randomly sampled data, e.g. data needed to assess the number of active cases
and various demographics of individuals that were affected. Census
provides an excellent  rationale. It is the only way one can develop rigorous estimates of various epidemiologically relevant quantities.

\noindent
There have been numerous surveys on the different types of datasets available for SARS-CoV-2, e.g.,~\cite{alamo2020open, alamo2020data, shuja2020covid,sameni2020mathematical}, as well as different kinds of modeling approaches. However, they do not describe how these models become relevant through the phases of pandemic response. An earlier similar attempt to summarize such response-driven modeling efforts can be found in \cite{wu2011use}, based on the 2009-H1N1 experience, this paper builds on their work and discusses these phases in the
present context and the SARS-CoV-2 pandemic. Although the paper touches upon different aspects of model-based decision making, we refer the readers to a companion article in the same special issue~\cite{adiga:jiisc20} for a focused review of models used for projection and forecasting.  

\section{Background}
\label{sec:background}
\begin{figure}
    \centering
    \includegraphics[width=0.75\textwidth]{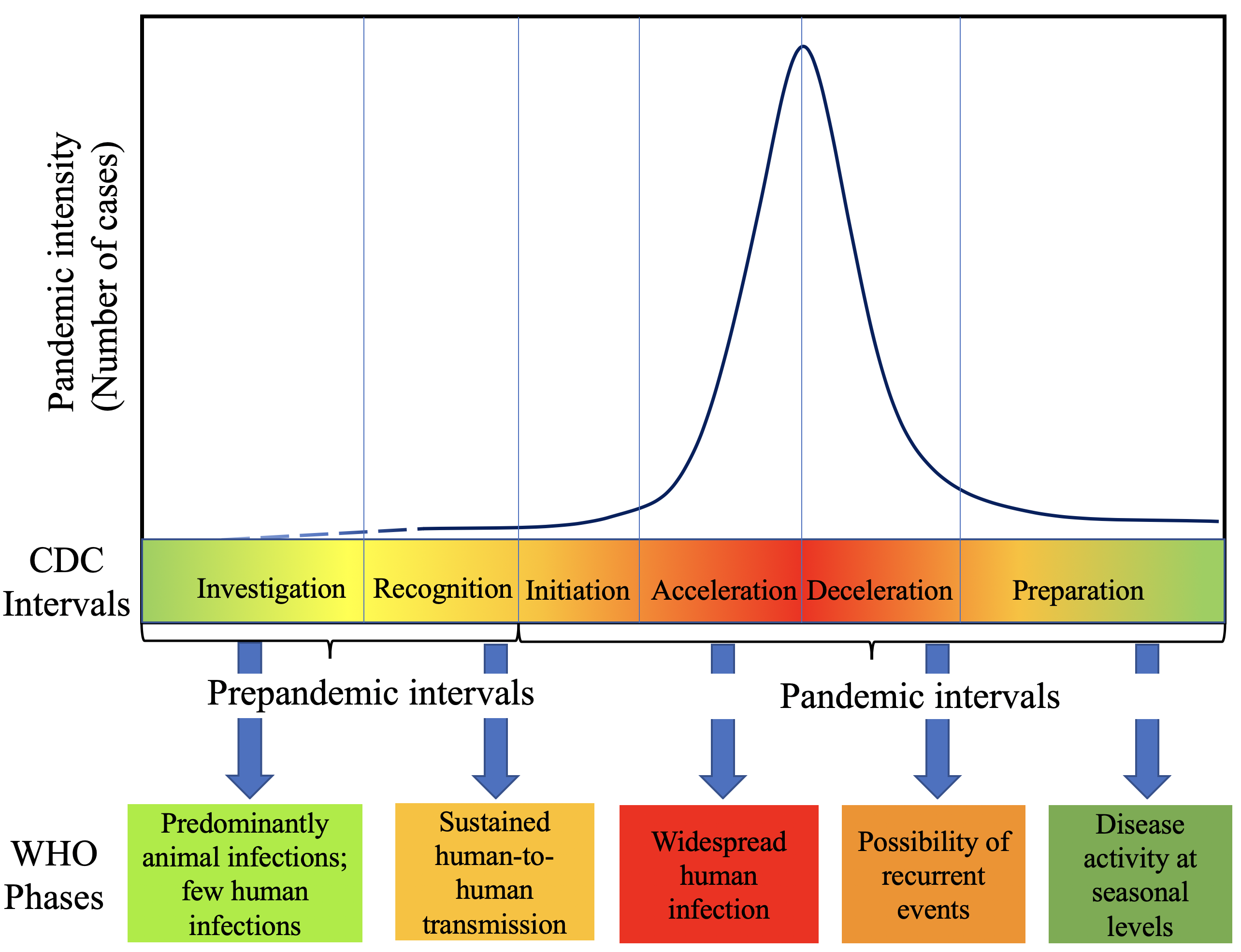}
    \caption{CDC Pandemic Intervals Framework and WHO phases for influenza pandemic}
\label{fig:pandemic_phases}
\end{figure}

\noindent
Multiple organizations including CDC and WHO have their frameworks for preparing and planning response to a pandemic. For instance, the Pandemic Intervals Framework from CDC\footnote{\url{https://www.cdc.gov/flu/pandemic-resources/national-strategy/intervals-framework.html}} describes the stages in the context of an influenza pandemic; these are illustrated in Figure~\ref{fig:pandemic_phases}. These six stages span investigation, recognition and initiation in the early phase, followed by most of the disease spread occurring during the acceleration and deceleration stages. They also provide indicators for identifying when the pandemic has progressed from one stage to the next \cite{holloway2014updated}. As envisioned, risk evaluation (i.e., using tools like Influenza Risk Assessment Tool (IRAT) and Pandemic Severity Assessment Framework (PSAF)) and early case identification characterize the first three stages, while non-pharmaceutical interventions (NPIs) and available therapeutics become central to the acceleration stage. The deceleration is facilitated by mass vaccination programs, exhaustion of susceptible population, or unsuitability of environmental conditions (such as weather). A similar framework is laid out in WHO's pandemic continuum\footnote{\url{https://www.cdc.gov/flu/pandemic-resources/planning-preparedness/global-planning-508.html}} and phases of pandemic alert\footnote{\url{https://www.who.int/csr/disease/swineflu/phase/en/}}. While such frameworks aid in streamlining the response efforts of these organizations, they also enable effective messaging. To the best of our knowledge, there has not been a similar characterization of mathematical modeling efforts that go hand in hand with supporting the response. 

\section{Modeling for stages of a pandemic}
\label{sec:stages}

\noindent
For summarizing the key models, we consider four of the stages of pandemic response  mentioned in Section~\ref{sec:background}: pre-pandemic, acceleration, mitigation and suppression. Here we provide the key problems in each stage, the datasets needed, the main tools and techniques used, and pertinent challenges. We structure our discussion based on our experience with modeling the spread of COVID-19 in the US, done in collaboration with local and federal agencies. 
\begin{itemize}
\item Pre-pandemic (Section~\ref{sec:prepandemic}): in the initial time period, there are few human infections, and the key questions involve understanding the epidemiological parameters, and the risks of importation to different countries. The primary sources of data used in this stage include line lists, clinical investigations and prior literature on similar diseases (for the former question), and mobility data such as airline flows, and information on travel restrictions.
\item Acceleration (Section~\ref{sec:acceleration}): 
this stage is relevant once the epidemic takes root within a country. There is usually a big lag in surveillance and response efforts, and the key questions are to model spread patterns at different spatio-temporal scales, and to derive short-term forecasts and projections. A broad class of datasets is used for developing models, including mobility, populations, land-use, and activities. These are combined with various kinds of time series data and covariates such as weather for forecasting. 
\item Mitigation (Section~\ref{sec:mitigation}): in this stage, different interventions, which are mostly non-pharmaceutical in the case of a novel pathogen, are implemented by government agencies, once the outbreak has taken hold within the population. This stage involves understanding the impact of interventions on case counts and health infrastructure demands, taking individual behaviors into account. The additional datasets needed in this stage include those on behavioral changes and hospital capacities. 
\item Suppression (Section~\ref{sec:suppression}): this stage involves designing methods to control the outbreak by contact tracing \& isolation and vaccination. Data on contact tracing, associated biases, vaccine production schedules, and compliance \& hesitancy are needed in this stage.  
\end{itemize}

\begin{figure}
    \centering
    \includegraphics[width=\textwidth]{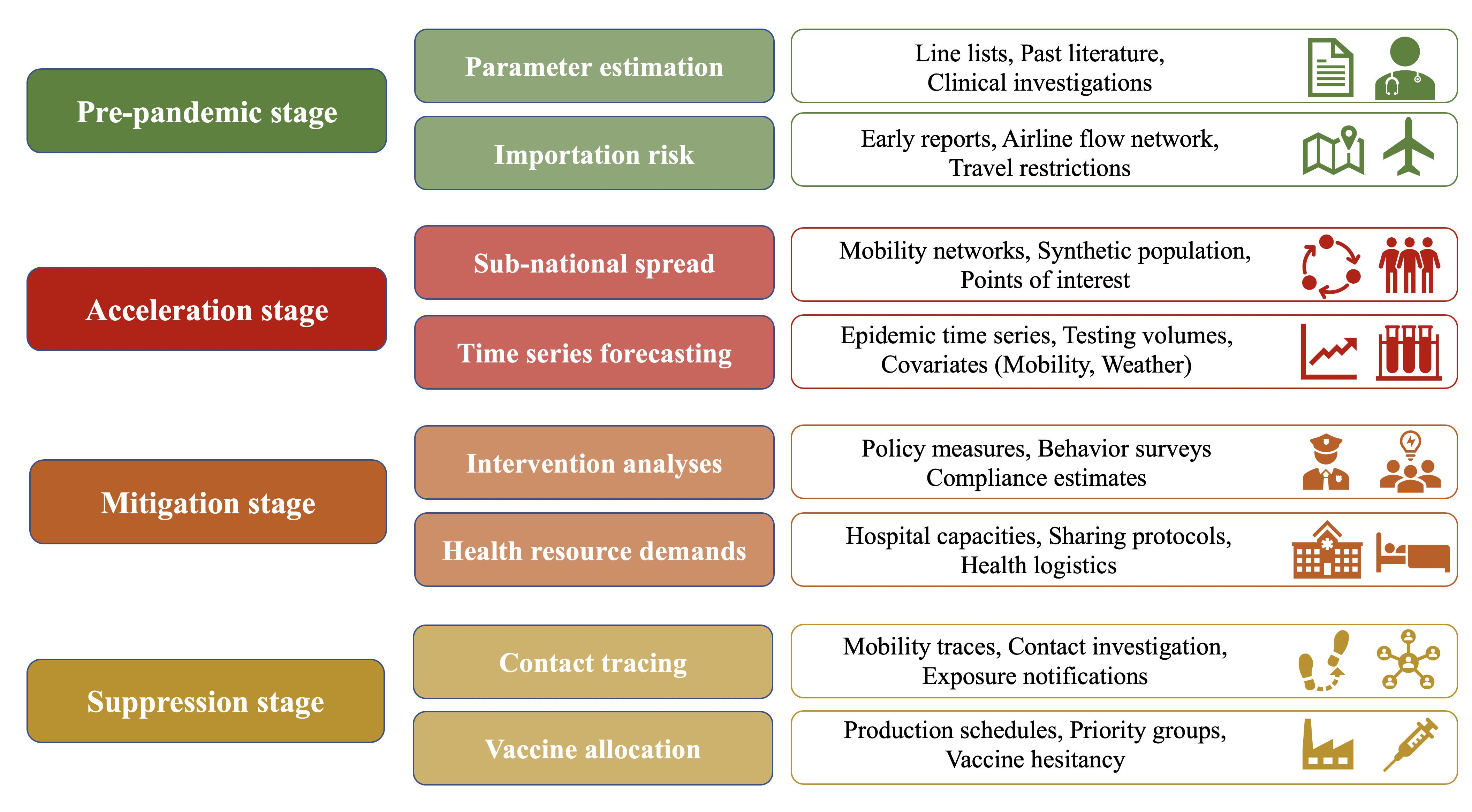}
    \caption{Summary of the data needs in different stages described in Section~\ref{sec:stages}.}
    \label{fig:stages-data}
\end{figure}

\noindent
Figure~\ref{fig:stages-data} gives an overview of this framework and summarizes the data needs in these stages. These stages also align well with the focus of the various modeling working groups organized by CDC which include epidemic parameter estimation, international spread risk, sub-national spread forecasting, impact of interventions, healthcare systems, and university modeling.
In reality, one should note that these stages may overlap, and may vary based on geographical factors and response efforts. Moreover, specific problems can be approached prospectively in earlier stages, or retrospectively during later stages. This framework is thus meant to be more conceptual than interpreted along a linear timeline. Results from such stages are very useful for policymakers to guide real-time response. 

\section{Pre-pandemic stage}
\label{sec:prepandemic}
\noindent
Consider a novel pathogen emerging in human populations that is detected through early cases involving unusual symptoms or unknown etiology. Such outbreaks are characterized by some kind of spillover event, mostly through zoonotic means, like in the case of COVID-19 or past influenza pandemics (e.g., swine flu and avian flu). A similar scenario can occur when an incidence of a well-documented disease with no known vaccine or therapeutics emerges in some part of the world, causing severe outcomes or fatalities (e.g., Ebola and Zika.) Regardless of the development status of the country where the pathogen emerged, such outbreaks now contains the risk of causing a worldwide pandemic due to the global connectivity induced by human travel.

Two questions become relevant at this stage: what are the epidemiological attributes of this disease, and what are the risks of importation to a different country? While the first question involves biological and clinical investigations, the latter is more related with societal and environmental factors. 

\subsection{Epidemiological parameter estimation}
\noindent
One of the crucial tasks during early disease investigation is to ascertain the transmission and severity of the disease. These are important dimensions along which the pandemic potential is characterized because together they determine the overall disease burden, as demonstrated within the Pandemic Severity Assessment Framework \cite{reed2013novel}. In addition to risk assessment for right-sizing response, they are integral to developing meaningful disease models.

\paragraph{Formulation}
Let $\Theta = \{\theta_T, \theta_S\}$ represent the transmission and severity parameters of interest. They can be further subdivided into sojourn time parameters $\theta_{\cdot}^{\delta}$ and transition probability parameters $\theta_{\cdot}^{p}$.  Here $\Theta$ corresponds to a continuous time Markov chain (CTMC) on the disease states. The problem formulation can be represented as follows:

Given $\Pi(\Theta)$, the prior distribution on the disease parameters and a dataset $\mathcal{D}$, estimate the posterior distribution $\mathbf{P}(\Theta | \mathcal{D})$ over all possible values of $\Theta$. In a model-specific form, this can be expressed as $\mathbf{P}(\Theta | \mathcal{D}, \mathcal{M})$ where $\mathcal{M}$ is a statistical, compartmental or agent-based disease model.

\begin{table}
\begin{center}
\begin{tabular}{|l|c|p{3.5in}|}
\hline
parameter & values & description \\
\hline
transmissibility ($R_0$) & 2.5 [2.0,3.0] & basic reproduction number\\
incubation period & 5 days & time from infection to onset\\
latent period & $3\sim 5$ days & time from infection to infectious\\
percent symptomatic & 65\% & infected people that exhibit symptoms\\
infectious period & $5\sim 6$ days & duration of infectiousness\\
infection detection rate & 15\% & 1 confirmed case corresponds to ~7 cases\\
serial interval & 7 days & time from infection to next generation infection\\
onset to hospitalization & 6.2 days & time from symptoms to hospitalization\\
hospitalization to ventilation & $1\sim 2$ days & time from hospitalization to ventilation\\
duration hospitalized & $3\sim 8$ days & time spent in the hospital\\
duration ventilated & $2\sim 7$ days & time spent on a ventilator\\
percent hospitalized & 5.5\% & symptomatic individuals becoming hospitalized\\
percent ventilated & 13\% & hospitalized patients that require ventilation\\
\hline
\end{tabular}
\end{center}
\caption{COVID-19 specific parameters that we currently use in our modeling and studies. Please note that the estimated values evolve in time; the values in the table are based on the {\em best guess 2020-04-14 version} of ``COVID-19 Pandemic Planning Scenarios'' document prepared by the Centers for Disease Control and Prevention (CDC) SARS-CoV-2 Modeling Team~\cite{cdc-model}.}
\label{tbl:disease-param}
\end{table}

\paragraph{Data needs}
In order to estimate the disease parameters sufficiently, line lists for individual confirmed cases is ideal. Such datasets contain, for each record, the date of confirmation, possible date of onset, severity (hospitalization/ICU) status, and date of recovery/discharge/death. Furthermore, age- and demographic/co-morbidity information allow development of models that are age- and risk group stratified. One such crowd-sourced line list was compiled during the early stages of COVID-19~\cite{xu2020epidemiological} and later released by CDC for US cases~\cite{COVID19C29:online}. Data from detailed clinical investigations from other countries such as China, South Korea, and Singapore was also used to parameterize these models~\cite{li2020covid}. In the absence of such datasets, past parameter estimates of similar diseases (e.g., SARS, MERS) were used for early analyses. 

\paragraph{Modeling approaches}
For a model agnostic approach, the delays and probabilities are obtained by various techniques, including Bayesian and Ordinary Least Squares fitting to various delay distributions. For a particular disease model, these are estimated through model calibration techniques such as MCMC and particle filtering approaches. A summary of community estimates of various disease parameters is provided at \url{https://github.com/midas-network/COVID-19}. Further such estimates allow the design of pandemic planning scenarios varying in levels of impact, as seen in the CDC scenarios page\footnote{\url{https://www.cdc.gov/coronavirus/2019-ncov/hcp/planning-scenarios.html}}.
See~\cite{ganyani2020estimating,lauer2020incubation,wu2020estimating} for methods and results related to estimating COVID-19 disease parameters from real data. Current models use a large set of disease parameters for modeling COVID-19 dynamics; they can be
broadly classified as transmission parameters and hospital
resource parameters. For instance in our work, we currently use parameters (with explanations) shown in 
Table~\ref{tbl:disease-param}.

\paragraph{Challenges}
Often these parameters are model specific, and hence one needs to be careful when reusing parameter estimates from literature. They are related but not identifiable with respect to population level measures such as basic reproductive number $R_0$ (or effective reproductive number $R_{\textrm{eff}}$) and doubling time which allow tracking the rate of epidemic growth. Also the estimation is hindered by inherent biases in case ascertainment rate, reporting delays and other gaps in the surveillance system. Aligning different data streams (e.g., outpatient surveillance, hospitalization rates, mortality records) is in itself challenging. 

\subsection{International importation risk}
\noindent
When a disease outbreak occurs in some part of the world, it is imperative for most countries to estimate their risk of importation through spatial proximity or international travel. Such measures are incredibly valuable in setting a timeline for preparation efforts, and initiating health checks at the borders. Over centuries, pandemics have spread faster and faster across the globe, making it all the more important to characterize this risk as early as possible.

\paragraph{Formulation}
Let $\mathcal{C}$ be the set of countries, and $\mathcal{G} = \{ \mathcal{C}, \mathcal{E} \}$ an international network, where edges (often weighted and directed) in $\mathcal{E}$ represent some notion of connectivity. The importation risk problem can be formulated as below:
\begin{displayquote}
Given $C_o \in \mathcal{C}$ the country of origin with an initial case at time $0$, and $C_i$ the country of interest, using $\mathcal{G}$, estimate the expected time taken $T_i$ for the first cases to arrive in country $C_i$.
\end{displayquote}
In its probabilistic form, the same can be expressed as estimating the probability $P_i(t)$ of seeing the first case in country $C_i$ by time $t$. 

\paragraph{Data needs}
Assuming we have initial case reports from the origin country, the first data needed is a network that connects the countries of the world to represent human travel. The most common source of such information is the airline network datasets, from sources such as IATA, OAG, and OpenFlights;~\cite{mesle2019use} provides a systematic review of how airline passenger data has been used for infectious disease modeling. These datasets could either capture static measures such as number of seats available or flight schedules, or a dynamic count of passengers per month along each itinerary. Since the latter has intrinsic delays in collection and reporting, for an ongoing pandemic they may not be representative. During such times, data on ongoing travel restrictions~\cite{bloomberg} become important to incorporate. Multi-modal traffic will also be important to incorporate for countries that share land borders or have heavy maritime traffic. For diseases such as Zika, where establishment risk is more relevant, data on vector abundance or prevailing weather conditions are appropriate.

\paragraph{Modeling approaches}
Simple structural measures on networks (such as degree, PageRank) could provide static indicators of vulnerability of countries. By transforming the weighted, directed edges into probabilities, one can use simple contagion models (e.g., Independent Cascades) to simulate disease spread and empirically estimate expected time of arrival. Global metapopulation models (GLEaM) that combine SEIR type dynamics with an airline network have also been used in the past for estimating importation risk. Brockmann and Helbing~\cite{brockmann:science13} used a similar framework to quantify effective distance on the network which seemed to be well correlated with time of arrival for multiple pandemics in the past; this has been extended to COVID-19~\cite{Adiga2020.02.20.20025882, chinazzi:science20}. In \cite{bogoch20}, the authors employ air travel volume obtained through IATA from ten major cities across China to rank various countries along with the IDVI to convey their vulnerability. \cite{Wu-nowcasting} consider the task of forecasting international and domestic spread of COVID-19 and employ Official Airline Group (OAG) data for determining air traffic to various countries, and \cite{salazar-underpred} fit a generalized linear model for observed number of cases in various countries as a function of air traffic volume obtained from OAG data to determine countries with potential risk of under-detection. Also,~\cite{gilbert-africa} provide Africa-specific case-study of vulnerability and preparedness using data from Civil Aviation Administration of China.

\paragraph{Challenges}
Note that arrival of an infected traveler will precede a local transmission event in a country. Hence the former is more appropriate to quantify in early stages. Also, the formulation is agnostic to whether it is the first infected arrival or first detected case. However, in real world, the former is difficult to observe, while the latter is influenced by security measures at ports of entry (land, sea, air) and the ease of identification for the pathogen. For instance, in the case of COVID-19, the long incubation period and the high likelihood of asymptomaticity could have resulted in many infected travelers being missed by health checks at PoEs. We also noticed potential administrative delays in reporting by multiple countries fearing travel restrictions.

\section{Acceleration stage}
\label{sec:acceleration}
As the epidemic takes root within a country, it may enter the acceleration phase. Depending on the testing infrastructure and agility of surveillance system, response efforts might lag or lead the rapid growth in case rate. Under such a scenario, two crucial questions emerge that pertain to how the disease may spread spatially/socially and how the case rate may grow over time. 

\subsection{Sub-national spread across scales}
Within the country, there is need to model the spatial spread of the disease at different scales: state, county, and community levels. Similar to the importation risk, such models may provide an estimate of when cases may emerge in different parts of the country. When coupled with vulnerability indicators (socio-economic, demographic, co-morbidities) they provide a framework for assessing the heterogeneous impact the disease may have across the country. Detailed agent-based models for urban centers may help identify hotspots and potential case clusters that may emerge (e.g., correctional facilities, nursing homes, food processing plants, etc. in the case of COVID-19).

\paragraph{Formulation}
Given a population representation $\mathcal{P}$ at appropriate scale and a disease model $\mathcal{M}$ per entity (individual or sub-region), model the disease spread under different assumptions of underlying connectivity $\mathcal{C}$ and disease parameters $\Theta$. The result will be a spatio-temporal spread model that results in $Z_{s,t}$, the time series of disease states over time for region $s$.

\paragraph{Data needs}
Some of the common datasets needed by most modeling approaches include: (1) social and spatial representation, which includes Census, and population data, which are available from Census departments (see, e.g.,~\cite{synthetic}), and Landscan~\cite{landscan}, (2) connectivity between regions (commuter, airline, road/rail/river), e.g.,~\cite{mesle2019use, bloomberg}, (3) data on locations, including points of interest, e.g., OpenStreetMap~\cite{openstreetmap}, and (4) activity data, e.g., the American Time Use Survey~\cite{atus}. These datasets help capture where people reside and how they move around, and come in contact with each other. While some of these are static, more dynamic measures, such as from GPS traces, become relevant as individuals change their behavior during a pandemic.

\paragraph{Modeling approaches}
Different kinds of structured metapopulation models~\cite{balcan:pnas2009, srini:ploscb19, gomes:plos-curr14, zhang:pnas17, chinazzi:science20}, and agent based models~\cite{ekmsw-2006,barrett:wsc09,eubank2004modelling, longini05:science,Ferg20} have been used in the past to model the sub-national spread; we refer to~\cite{Brauer-2008,Ne03, marathe:cacm13} for surveys on different modeling approaches. These models incorporate typical mixing patterns, which result from detailed activities and co-location (in the case of agent based models), and different modes of travel and commuting (in the case of metapopulation models).

\paragraph{Challenges}
While metapopulation models can be built relatively rapidly, agent based models are much harder---the datasets need to be assembled at a large scale, with detailed construction pipelines, see, e.g.,~\cite{ekmsw-2006,barrett:wsc09,eubank2004modelling, longini05:science,Ferg20}. Since detailed individual activities drive the dynamics in agent based models, schools and workplaces have to be modeled, in order to make predictions meaningful. Such models will get reused at different stages of the outbreak, so they need to be generic enough to incorporate dynamically evolving disease information. Finally, a common challenge across modeling paradigms is the ability to calibrate to the dynamically evolving spatio-temporal data from the outbreak---this is especially challenging in the presence of reporting biases and data insufficiency issues. 

\subsection{Growth rate and time series forecasting}
Given the early growth of cases within the country (or sub-region), there is need for quantifying the rate of increase in comparable terms across the duration of the outbreak (accounting for the exponential nature of such processes). These estimates also serve as references, when evaluating the impact of various interventions. As an extension, such methods and more sophisticated time series methods can be used to produce short-term forecasts for disease evolution. 

\paragraph{Formulation}
Given the disease time series data within the country $Z_{s,t}$ until data horizon $T$, provide scale-independent growth rate measures $G_s(T)$, and forecasts $\hat{Z}_{s,u}$ for $u \in [T, T+\Delta T]$, where $\Delta T$ is the forecast horizon. 

\paragraph{Data needs}
Models at this stage require datasets such as (1) time series data on different kinds of disease outcomes, including case counts, mortality, hospitalizations, along with attributes, such as age, gender and location, e.g.,~\cite{awsdata, midasdata, nytimesdataset, umd, coviddashboard}, (2) any associated data for reporting bias (total tests, test positivity rate)~\cite{covidtracking}, which need to be incorporated into the models, as these biases can have a significant impact on the dynamics, and (3) exogenous regressors (mobility, weather), which have been shown to have a significant impact on other diseases, such as Influenza, e.g.,~\cite{shaman:plos10}. 

\paragraph{Modeling approaches}
Even before building statistical or mechanistic time series forecasting methods, one can derive insights through analytical measures of the time series data. For instance, the effective Reproductive number, estimated from the time series~\cite{cori2013epiestim} can serve as a scale-independent metric to compare the outbreaks across space and time. Additionally multiple statistical methods ranging from autoregressive models to deep learning techniques can be applied to the time series data, with additional exogenous variables as input. While such methods perform reasonably for short-term targets, mechanistic approaches as described earlier can provide better long-term projections. Various ensembling techniques have also been developed in the recent past to combine such multi-model forecasts to provide a single robust forecast with better uncertainty quantification. One such effort that combines more than 30 methods for COVID-19 can be found at the COVID Forecasting Hub\footnote{\url{https://covid19forecasthub.org/}}. We also point to the companion paper for more details on projection and forecasting models. 

\paragraph{Challenges}
Data on epidemic outcomes usually has a lot of uncertainties and errors, including missing data, collection bias, and backfill. For forecasting tasks, these time series data need to be near real-time, else one needs to do both nowcasting, as well as forecasting. Other exogenous regressors can provide valuable lead time, due to inherent delays in disease dynamics from exposure to case identification. Such frameworks need to be generalized to accommodate qualitative inputs on future policies (shutdowns, mask mandates, etc.), as well as behaviors, as we discuss in the next section.

\section{Mitigation stage}
\label{sec:mitigation}

Once the outbreak has taken hold within the population, local, state and national governments attempt to mitigate and control its spread by considering different kinds of interventions.
Unfortunately, as the COVID-19 pandemic has shown, there is a significant delay in the time taken by governments to respond. As a result, this has caused a large number of cases, a fraction of which lead to hospitalizations. Two key questions in this stage are: (1) how to evaluate different kinds of interventions, and choose the most effective ones, and (2) how to estimate the healthcare infrastructure demand, and how to mitigate it. The effectiveness of an intervention (e.g., social distancing) depends on how individuals respond to them, and the level of compliance. The health resource demand depends on the specific interventions which are implemented. As a result, both these questions are connected, and require models which incorporate appropriate behavioral responses.

\subsection{Intervention analyses}

In the initial stages, only non-prophylactic interventions are available, such as: social distancing, school and workplace closures, and use of PPEs, since
no vaccinations and anti-virals are available. As mentioned above, such analyses are almost entirely model based, and the specific model depends on the nature of the intervention and the population being studied.

\paragraph{Formulation}
Given a model, denoted abstractly as $\mathcal{M}$, the general goals are (1) to evaluate the impact of an intervention (e.g., school and workplace closure, and other social distancing strategies) on different epidemic outcomes (e.g., average outbreak size, peak size, and time to peak), and (2) find the most effective intervention from a suite of interventions, with given resource constraints. The specific formulation depends crucially on the model and type of intervention. Even for a single intervention, evaluating its impact is quite challenging, since there are a number of sources of uncertainty, and a number of parameters associated with the intervention (e.g., when to start school closure, how long, and how to restart). Therefore, finding uncertainty bounds is a key part of the problem.

\paragraph{Data needs}
While all the data needs from the previous stages for developing a model are still there, representation of different kinds of behaviors is a crucial component of the models in this stage; this includes: use of PPEs, compliance to social distancing measures, and level of mobility.
Statistics on such behaviors are available at a fairly detailed level (e.g., counties and daily) from multiple sources, such as (1) the COVID-19 Impact Analysis Platform from the University of Maryland~\cite{umd}, which gives metrics related to social distancing activities, including level of staying home, outside county trips, outside state trips, (2) changes in mobility associated with different kinds of activities from Google~\cite{google-mobility}, and other sources, (3) survey data on different kinds of behaviors, such as usage of masks~\cite{nytimes-mask}.

\paragraph{Modeling approaches}
As mentioned above, such analyses are almost entirely model based, including structured metapopulation models~\cite{balcan:pnas2009, srini:ploscb19, gomes:plos-curr14, zhang:pnas17, chinazzi:science20}, and agent based models~\cite{ekmsw-2006,barrett:wsc09,eubank2004modelling, longini05:science,Ferg20}.
Different kinds of behaviors relevant to such interventions, including compliance with using PPEs and compliance to social distancing guidelines, need to be incorporated into these models. Since there is a great deal of heterogeneity in such behaviors, it is conceptually easiest to incorporate them into agent based models, since individual agents are represented. However, calibration, simulation and analysis of such models pose significant computational challenges. On the other hand, the simulation of metapopulation models is much easier, but such behaviors cannot be directly represented--- instead, modelers have to estimate the effect of different behaviors on the disease model parameters, which can pose modeling challenges.

\paragraph{Challenges}
There are a number of challenges in using data on behaviors, which depends on the specific datasets. Much of the data available for COVID-19 is estimated through indirect sources, e.g., through cell phone and online activities, and crowd-sourced platforms. This can provide large spatio-temporal datasets, but have unknown biases and uncertainties. On the other hand, survey data is often more reliable, and provides several covariates, but is typically very sparse. Handling such uncertainties, rigorous sensitivity analysis, and incorporating the uncertainties into the analysis of the simulation outputs are important steps for modelers.

\subsection{Health resource demands}

The COVID-19 pandemic has led to a significant increase in hospitalizations. Hospitals are typically optimized to run near capacity, so there have been fears that the hospital capacities would not be adequate, especially in several countries in Asia, but also in some regions in the US. Nosocomial transmission could further increase this burden.

\paragraph{Formulation}
The overall problem is to estimate the demand for hospital resources within a population---this includes the number of hospitalizations, and more refined types of resources, such as ICUs, CCUs, medical personnel and equipment, such as ventilators. An important issue is whether the capacity of hospitals within the region would be overrun by the demand, when this is expected to happen, and how to design strategies to meet the demand---this could be through augmenting the capacities at existing hospitals, or building new facilities. Timing is of essence, and projections of when the demands exceed capacity are important for governments to plan.

\paragraph{Data needs}
The demands for hospitalization and other health resources can be estimated from the epidemic models mentioned earlier, by incorporating suitable health states, e.g.,~\cite{wang:eid20, srini:ploscb19}; in addition to the inputs needed for setting up the models for case counts, datasets are needed for hospitalization rates and durations of hospital stay, ICU care, and ventilation.
The other important inputs for this component are hospital capacity, and the referral regions (which represent where patients travel for hospitalization). Different public and commercial datasets provide such information, e.g.,~\cite{cdc-hospital, hosp2}. 

\paragraph{Modeling approaches}
Demand for health resources is typically incorporated into both metapopulation and agent based models, by having a fraction of the infectious individuals transition into a hospitalization state. An important issue to consider is what happens if there is a shortage of hospital capacity. Studying this requires modeling the hospital infrastructure, i.e., different kinds of hospitals within the region, and which hospital a patient goes to. There is typically limited data on this, and data on hospital referral regions, or voronoi tesselation can be used. Understanding the regimes in which hospital demand exceeds capacity is an important question to study. Nosocomial transmission is typically much harder to study, since it requires more detailed modeling of processes within hospitals.

\paragraph{Challenges}
There is a lot of uncertainty and variability in all the datasets involved in this process, making its modeling difficult. For instance, forecasts of the number of cases and hospitalizations have huge uncertainty bounds for medium or long term horizon, which is the kind of input necessary for understanding hospital demands, and whether there would be any deficits. 

\section{Suppression stage}
\label{sec:suppression}
The suppression stage involves methods to control the outbreak, including reducing the incidence rate and potentially leading to the eradication of the disease in the end. Eradication in case of COVID-19 appears unlikely as of now, what is more likely is that this will become part of seasonal human coronaviruses that will mutate continuously much like the influenza virus. 

\subsection{Contact tracing, testing and isolation}
Contact tracing problem refers to the ability to trace the neighbors of
an infected individual. Ideally, if one is successful, each neighbor of an infected neighbor would be identified and isolated from the larger population to reduce the growth of a pandemic. In some cases, each such neighbor could be tested to see if the individual has contracted the disease. Contact tracing is the workhorse in epidemiology and has been immensely successful in controlling slow moving diseases. When combined with vaccination and other pharmaceutical interventions, it provides the best way to control and suppress an epidemic.

\paragraph{Formulation}
The basic contact tracing problem is stated as follows: Given a social contact network $G(V,E)$ and subset of nodes $S \subset V$ that are infected and a subset $S_1 \subset S$ of nodes identified as infected,
find all neighbors of $S$. Here a neighbor means an individual
who is likely to have a substantial contact with the infected person.
One then tests them (if tests are available), and following that, isolates these neighbors, or vaccinates them or administers anti-viral.
The measures of effectiveness for the problem include: ($i$) maximizing the size of $S_1$, ($ii$) maximizing the size of set $N(S_1) \subseteq N(S)$, i.e. the potential number of neighbors of set $S_1$,
($iii$) doing this within a short period of time so that these neighbors either do not become infectious, or they minimize the number of days that they are infectious, while they are still interacting in the community in a normal manner, ($iv$) 
the eventual goal is to try and reduce the incidence rate in the community---thus if all the neighbors of $S_1$ cannot be identified, one aims to identify those individuals who when isolated/treated lead to a large impact;
($v$) and finally verifying that these individuals indeed came in contact with the infected individuals and thus can be asked to isolate or be treated.
\paragraph{Data needs}
Data needed for the contact tracing problem includes: ($i$) a line list of individuals who are currently known to be infected (this is needed in case
of human based contact tracing). In the real world, when carrying out human contact tracers based deployment, one interviews all the individuals who are known to be infectious and reaches out to their contacts.
\paragraph{Modeling approaches}
Human contact tracing is routinely done in epidemiology. Most states in the US have hired such contact tracers. They obtain the daily incidence report from the state health departments and then proceed to contact the individuals who are confirmed to be infected. Earlier, human contact tracers used to go from house to house and identify the potential neighbors through a well defined interview process.  Although very effective it is very time consuming and labor intensive. Phones were used extensively in the last 10-20 years as they allow the contact tracers to reach individuals. They are helpful but have the downside that it might be hard to reach all individuals. During COVID-19 outbreak, for the first time,
societies and governments have considered and deployed digital contact tracing tools~\cite{lorch2020spatiotemporal,salathe2020covid,ferretti2020quantifying,kretzschmar2020isolation,chan2020pact}. These can be quite effective but also have certain weaknesses, including, privacy, accuracy, and limited market penetration of the digital apps.

\paragraph{Challenges}
These include:
($i$) inability to identify everyone who is infectious (the set $S$) --- this is virtually impossible for COVID-19 like disease unless the incidence rate has come down drastically and for the reason that many individuals are infected but asymptomatic; ($ii$) identifying all contacts of $S$ (or $S_1$) -- this is hard since individuals cannot recall everyone they met,
certain folks that they were in close proximity might have been in stores or social events and thus not known to individuals in the set $S$. Furthermore, even if a person is able to identify the contacts, it is often hard to reach all the individuals due to resource constraints (each human tracer can only contact a small number of individuals.

\subsection{Vaccine allocation}
The overall goal of the vaccine allocation problem is to allocate vaccine
efficiently and in a timely manner to reduce the overall burden of the pandemic.
\paragraph{Formulation}
The basic version of the problem can be cast in a very simple manner
(for networked models): Given a graph $G(V,E)$ and a budget $B$
on the number of vaccines available, find a set $S$ of size $B$ to vaccinate so as to optimize certain measure of effectiveness. The measure of effectiveness can be ($i$) minimizing the total number of individuals
infected (or maximizing the total number of uninfected individuals); 
($ii$) minimizing the total number of deaths
(or maximizing the total number of deaths averted); ($iii$) optimizing the above quantities but keeping in mind certain equity and fairness criteria
(across socio-demographic groups, e.g. age, race, income);
($iv$) taking into account vaccine hesitancy of individuals;
($v$) taking into account the fact that all vaccines are not available
at the start of the pandemic, and when they become available, one gets
limited number of doses each month; ($vi$) deciding how to share the stockpile between countries, state, and other organizations; ($vii$)
taking into account efficacy of the vaccine.

\paragraph{Data needs}
As in other problems, vaccine allocation problems need as input a good representation of the system; network based, meta-population based and compartmental mass action models can be used. One other key input is the vaccine budget, i.e., the production schedule and timeline, which serves as the constraint for the allocation problem. Additional data on prevailing vaccine sentiment and past compliance to seasonal/neonatal vaccinations are useful to estimate coverage.

\paragraph{Modeling approaches}
The problem has been studied actively in the literature; network science
community has focused on optimal allocation schemes, while public health community has focused on using meta-population models and assessing 
certain fixed allocation schemes based on socio-economic and demographic
considerations.  Game theoretic
approaches that try and understand strategic behavior of individuals
and organization has also been studied.
\paragraph{Challenges}
The problem is computationally challenging and thus most of the time simulation based optimization techniques are used. Challenge to the optimization approach comes from the fact that the optimal allocation
scheme might be hard to compute or hard to implement. Other challenges include  fairness criteria (e.g. the optimal set might be a specific group) and also multiple objectives that one needs to balance.

\section{Discussion}
While the above sections provide an overview of salient modeling questions that arise during the key stages of a pandemic, mathematical and computational model development is equally if not more important as we approach the post-pandemic (or more appropriately \emph{inter-pandemic}) phase. Often referred to as \emph{peace time} efforts, this phase allows modelers to retrospectively assess individual and collective models on how they performed during the pandemic. In order to encourage continued development and identifying data gaps, synthetic forecasting challenge exercises \cite{viboud2018rapidd} may be conducted where multiple modeling groups are invited to forecast synthetic scenarios with varying levels of data availability. Another set of models that are quite relevant for policymakers during the winding down stages, are those that help assess overall health burden and economic costs of the pandemic. 
\medskip

\noindent
\textbf{Acknowledgments.} 
The authors would like to thank members of the Biocomplexity COVID-19 Response Team and Network Systems Science and Advanced Computing (NSSAC) Division for their thoughtful comments and suggestions related to epidemic modeling and response support. We thank members of the Biocomplexity Institute and Initiative, University of Virginia for useful discussion and suggestions. 
This work was partially supported by National Institutes of Health (NIH) Grant R01GM109718, NSF BIG DATA Grant IIS-1633028, NSF DIBBS Grant OAC-1443054, 
NSF Grant No.: OAC-1916805, NSF Expeditions in Computing Grant CCF-1918656, CCF-1917819, NSF RAPID CNS-2028004, NSF RAPID OAC-2027541,  US Centers for Disease Control and Prevention 75D30119C05935, DTRA subcontract/ARA S-D00189-15-TO-01-UVA. Any opinions, findings, and conclusions or recommendations expressed in this material are those of the author(s) and do not necessarily reflect the views of the funding agencies.


\bibliography{refs.bib}
\bibliographystyle{unsrt}
\newpage

\end{document}